\documentclass[12pt]{article}
\usepackage{amssymb}

\newcommand{\nc}{\newcommand}
\nc{\la}{\lambda} \nc{\alf}{\alpha}
\nc{\tht}{\theta}  \nc{\be}{\beta}  \nc{\eps}{\epsilon}
\nc{\ga}{\gamma}  \nc{\De}{\Delta}  \nc{\Ga}{\Gamma}  \nc{\vphi}{\varphi}
\nc{\de}{\delta} \nc{\si}{\sigma}  \nc{\ka}{\kappa}  \nc{\T}{\Theta}
\nc{\om}{\omega}  \nc{\Om}{\Omega}  
\nc{\qq}{\quad\quad}  \nc{\nf}{\infty}   \nc{\dl}{\mathop{\smash{\cal L}}}
\nc{\ra}{\rightarrow}  \nc{\ol}{\overline}  \nc{\und}{\underline}
\nc{\beq}{\begin{equation}}   \nc{\eeq}{\end{equation}}
\nc{\beqa}{\begin{eqnarray}}  \nc{\eeqa}{\end{eqnarray}}
\nc{\nin}{\noindent}          \nc{\pt}{\partial}
\nc{\dst}{\displaystyle}      \nc{\nnb}{\nonumber}
\nc{\bs}{\backslash}          \nc{\mb}{\mathbb} 
\nc{\dg}{\dagger}   \nc{\wh}{\widehat}  \nc{\wti}{\widetilde}
\nc{\PT}{\Pi_{\tht}}  \nc{\PF}{\Pi_{\phi}}       \nc{\R}{\mb R}

\newcounter{muni}
\newenvironment{remunerate}{\begin{list}{{\rm \arabic{muni}.}}
{\usecounter{muni}
\setlength{\leftmargin}{0pt}\setlength{\itemindent}{38pt}}}{\end{list}}

\nc{\brm}{\begin{remunerate}}   \nc{\erm}{\end{remunerate}}
\nc{\barr}{\begin{array}}   \nc{\earr}{\end{array}}

\nc{\stg}{\mathop{\smash{*}}}    \nc{\st}{\mathop{\smash{\delta}}}
 
\nc{\mtvb}{\mathversion{bold}}   \nc{\mtvn}{\mathversion{normal}}

\begin{document}

\begin{titlepage}

\[  \]
\centerline{\Large\bf  Bianchi type II, III and V }

\vspace{4mm}
\centerline{\Large\bf Diagonal Einstein metrics re-visited}

\vskip 2.0truecm
\centerline{\large\bf Galliano VALENT}

\vskip 2.0truecm
\centerline{ \it Laboratoire de Physique Th\'eorique et des
Hautes Energies}
\centerline{\it CNRS, UMR 7589}
\centerline{\it 2 Place Jussieu, F-75251 Paris Cedex 05, France}
\nopagebreak

\vskip 2.5truecm

\begin{abstract}
We present, for both minkowskian and euclidean signatures, short derivations of the 
diagonal Einstein metrics for Bianchi type II, III and V. For the first two cases we show 
the integrability of the geodesic flow while for the third case a somewhat unusual 
bifurcation phenomenon takes place: for minkowskian signature elliptic functions are essential 
in the metric while for euclidean signature only elementary functions appear.

\end{abstract}

\end{titlepage}

\section{Introduction}
Modern cosmology \cite{we} has led to a strong development of models based on Bianchi cohomogeneity 
one metrics. A large amount of information was gathered, mainly for Ricci-flat Bianchi type A metrics (see \cite{sk}) either in minkowskian or in euclidean signature. However the need for a cosmological constant leads to consider rather Einstein metrics and not just Ricci-flat ones, leading to more difficult problems. 

For the minkowskian signature a complete list of the 
algebraically special and hyper\-surface-homogeneous Einstein metrics, using spinors, 
is given in \cite{Mc} and many others appear in \cite{mm}. For the euclidean signature 
the most impressive progresses came from Weyl tensor self-duality and culminated with the 
tri-axial Bianchi type IX self-dual Einstein metrics of Tod and Hitchin \cite{To}, \cite{Hi}. However these ideas give only limited results for the type B metrics as observed in \cite{Tobis}. Another difficulty linked to the type B studies is that even for Ricci-flat geometries 
there is no need for the metric to be diagonal in the invariant one-forms. Despite these difficulties, quite recently new results were derived for type III \cite{ct}, type V \cite{cg} 
and type VII$_h$ \cite{tc} for which the most general vacuum minkowskian metrics, i. e. 
non-diagonal ones, were derived.

Our aim is to give very simple derivations of some Einstein 
metrics, for both euclidean and minkowskian signatures, under the simplifying hypothesis 
that the metric is {\em diagonal} with respect to the invariant spatial one-forms. These metrics 
are examined for Bianchi metrics of type II, III and V. As we shall see they both 
exhibit interesting features: the types II and III have an integrable geodesic flow and the type V 
presents an interesting ``bifurcation" between the minkowskian and the euclidean regime. 

As pointed out by a Referee, the basic integrability of the Einstein equations in the 
cases considered in this article, which results in our work from an appropriate fixing of 
the time coordinate, is best understood in a unified and systematic approach if one uses the 
hamiltonian formalism developed by Uggla et al in \cite{ujr}. They have shown that the 
existence of Killing tensors is a key tool leading to a sytematic display of the cases 
leading to integrability, even if one considers matter and not merely a cosmological constant. 
This approach is in some sense reminiscent of Carter's derivation of Kerr metric by 
imposing that it must have a Killing tensor \cite{Ca}.

The content of this article is the following: in Section 2 we present background informations 
and the field equations for the bi-axial Bianchi type II metrics (they have one extra Killing vector). The metrics are then constructed and, being of type D, their geodesic flow is shown 
to be integrable. 

In Section 3 we present the corresponding construction for the type III metrics. They all 
share one extra Killing vector. All these metrics are of type D with an integrable 
geodesic flow. However, in some special cases, there is a strong symmetry enhancment 
leading to de Sitter, anti de Sitter and ${\mb H}^4$ in somewhat unusual coordinates

In section 4 we present the corresponding construction for the type V metrics. Special 
cases include again de Sitter, anti de Sitter and ${\mb H}^4$. However in general one needs 
elliptic functions to express the minkowskian metrics whereas for the euclidean ones only 
elementary functions appear.

We give in Appendix A more details on the curious forms of de Sitter metric encountered 
in the analysis of the Bianchi type III and V Einstein metrics, in Appendix B some 
technicalities related to elliptic functions and in Appendix C some checks involving 
curvature computations.

\section{Type II metrics}
The Bianchi type II Lie algebra is defined as
\beq\label{2lie}
[{\cal L}_1,{\cal L}_2]=0,\qq [{\cal L}_2,{\cal L}_3]={\cal L}_1,\qq [{\cal L}_3,{\cal L}_1]=0.\eeq
One can choose ``spatial" coordinates $(x,y,z)$ such that
\beq\label{2iso}
{\cal L}_1=\pt_x,\qq {\cal L}_2=\pt_y-z\,\pt_x,\qq {\cal L}_3=\pt_z,\eeq
and the invariant 1-forms
\beq\label{2mc}
\si_1=dx+ydz,\qq \si_2=dy,\qq \si_3=dz.\eeq
We will look for diagonal metrics of the form
\beq\label{metb}
g=\beta^2\,\si_1^2+\ga^2(\si_2^2+\si_3^2)+\eps\,\alpha^2\,dt^2.
\eeq
The bi-axial character of this metric gives a fourth Killing vector
\beq
{\cal L}_4=y\,\pt_z-z\,\pt_y-\frac 12(y^2-z^2)\pt_x,\eeq
and the algebra closes according to
\beq
[{\cal L}_4,{\cal L}_2]=-{\cal L}_3,\qq [{\cal L}_4,{\cal L}_3]={\cal L}_2.\eeq 

\subsection{Integration of the field equations}
The Einstein equations \footnote{In our notations the spheres have positive curvature.}
\[Ric_{~\mu}^{~\nu}=\la\,\de_{~\mu}^{~\nu}\] 
give 4 independent equations
\[\barr{ll}   
(I)\quad & \dst\frac{\ddot{\be}}{\be}+\frac{\dot{\be}}{\be}\left(
2\frac{\dot{\ga}}{\ga}-\frac{\dot{\alf}}{\alf}\right)-\eps\frac{\alf^2\be^2}{2\ga^4}
+\eps\la\alf^2=0,\\[5mm]
(II)\quad & \dst \frac{\ddot{\ga}}{\ga}+\frac{\dot{\ga}}{\ga}\left(\frac{\dot{\be}}{\be}
+\frac{\dot{\ga}}{\ga}-\frac{\dot{\alf}}{\alf}\right)+\eps\frac{\alf^2\be^2}{2\ga^4}+\eps\,\la\,\alf^2=0, \\[5mm]    
(III)\quad & \dst \frac{\ddot{\be}}{\be}+2\frac{\ddot{\ga}}{\ga}-\frac{\dot{\alf}}{\alf}
\left(\frac{\dot{\be}}{\be}+2\frac{\dot{\ga}}{\ga}\right)+\eps\la\alf^2=0.\earr\]
The last relation, using the first two, simplifies to
\[(IV)\qq 4\frac{\dot{\be}\dot{\ga}}{\be\ga}+2\frac{\dot{\ga}^2}{\ga^2}
+\eps\frac{\alf^2\be^2}{2\ga^4}+2\eps\la\alf^2=0.\]
Subtracting (IV) to twice (II) we get
\beq
2\frac{\ddot{\ga}}{\ga}-2\frac{\dot{\ga}}{\ga}\left(
\frac{\dot{\be}}{\be}+\frac{\dot{\alf}}{\alf}\right)+\eps\frac{\alf^2\be^2}{2\ga^4}=0,\eeq
which suggest to fix up the time coordinate arbitrariness by imposing $\,\alf\,\be=1$. The previous relation decouples to the integrable equation
\beq\label{2eq1}
\frac{\ddot{\ga}}{\ga}+\frac{\eps}{4\ga^4}=0\qq\Longrightarrow\qq
\dot{\ga}^2-\frac{\eps}{4\ga^2}=E.\eeq
Defining $\,\rho=\be^2\ga^2,$ and combining (I) with (II) we get
\beq\label{2eq2}
\ddot{\rho}=-4\eps\la\,\ga^2,\eeq
and relation (IV) reduces to
\beq\label{2eq3}
\frac{\dot{\rho}}{\rho}\frac{\dot{(\ga^2)}}{\ga^2}+\frac{\eps}{2\ga^4}
+2\eps\la\,\frac{\ga^2}{\rho}=0.
\eeq
So we need to integrate (\ref{2eq1}) for $\ga$, then compute $\rho$ and impose (\ref{2eq3}). Let us discuss separately the two signatures.

\subsection{Minkowskian signature}
In this case $E$ cannot vanish. One gets
\beq
\ga^2=E\left(t^2+\frac 1{4E^2}\right),\qq \rho=\rho_0+mt
+\la E\left(\frac{t^2}{2E^2}+\frac{t^4}{3}\right).\eeq
Imposing (\ref{2eq3}) we get $\dst\rho_0=-\frac{\la}{16E^3}$ and $m$ remains free. In order to get rid of the factor $E$ in $\la E$ it is sufficient to divide the metric by $E$. To compare to previous work let us define $\,4l^2E^2=1$. After obvious algebra we end up with
\beq\label{2mE}\left\{\barr{l}\dst 
g=4l^2\frac uc\,\si_1^2-\frac cu\,dt^2+c(\si_2^2+\si_3^2),\\[4mm]\dst 
c=t^2+l^2,\qq u=mt+\la\left(-l^4+2l^2t^2+\frac{t^4}{3}\right).\earr\right.\eeq
First obtained by Cahen and Defrise \cite{cd}, see formula (13.48) with $e=k=0$ in \cite{sk}. It 
has Petrov type D.

\subsection{Euclidean signature}
In this case $E=0$ is possible. Let us first dispose with this case. Using $t=\ga^2$ as a new variable we have $\,\rho=l+mt-\frac 23\la\,t^3$. This time (\ref{2eq3}) requires $m=0$. 
So the metric can be written
\beq
g=\Delta\,\si_1^2+\frac{dt^2}{\Delta}+t(\si_2^2+\si_3^2),\qq\Delta=\frac lt-\frac{2\la}{3}t^2.\eeq
This metric was obtained by Dancer and Strachan \cite{ds}. It is K\"ahler, with complex 
structure $J=dt\wedge\si_1+t\,\si_2\wedge\si_3$.

For $E\neq 0$ there are 2 cases, according to its sign. Since the derivations are rather 
similar to the minkowskian case, let us just state the results. For $E>0$, with the same 
relation  between $l$ and $E$, we obtain
\beq\label{lo}\left\{\barr{l}\dst 
g=4l^2\frac uc\,\si_1^2+\frac cu\,dt^2+c(\si_2^2+\si_3^2),\\[4mm]\dst
c=t^2-l^2,\qq u=mt+\la\left(l^4+2l^2t^2-\frac{t^4}{3}\right).\earr\right.\eeq
The metric for $E<0$ follows from (\ref{lo}) by the changes $\,c\to -c$ and $\la\to-\la$.

Let us first observe that the parameter $l$ is not essential: one can get rid of it by the changes
\[g\to \frac g{l^4},\qq\tau=\frac tl,\qq x\to \frac x{l^2},\qq y\to \frac yl,\qq z\to \frac zl.\]
Therefore the metric displays two essential constants: $m$ and $\la$ which are expected for the general solution.

The metric (\ref{lo}) was first derived by Lorenz-Petzold in \cite{Lo}. The Weyl tensor is 
\beq
W^+=\frac{3m+8\la l^3}{6(t-l)^3}\,M,\qq W^-=\frac{3m-8\la l^3}{3(t+l)^3}\,M,\quad 
M={\rm diag}\,(-2,1,1),\eeq
so there is room for metrics with self-dual Weyl tensor. Let us consider the case 
\beq
W^+=0 \quad\to\quad u=-\frac{\la}{3}(t+3l)(t-l)^3,\qq t\geq l.\eeq
Positivity requires $\la<0$, so that  defining 
\[ \frac sl=\frac{t-2b}{t+2b},\qq\quad -8b=\frac 3{2|\la|l^2},\]
the metric becomes
\beq
g_{SD}=\frac 3{2|\la|(t+2b)^2}\left(\frac{t+b}{t}\,\si_1^2
+\frac t{t+b}\,dt^2+t(\si_2^2+\si_3^2)\right),
\eeq
which was derived in \cite{vy}, and shown to be complete for $b=-1$.

\subsection{Integrable geodesic flow}
It is known \cite{wp}, \cite{Co} that most Petrov type D vacuum metrics (the C metric 
being a notable exception) exhibit at least one Killing-Yano tensor, the square of which induces a Killing-St\"ackel tensor (following the same terminology as in \cite{vy}) and this last one 
is essential to the integrability of the geodesic flow. However we suspect that this 
property could remain true for many type D Einstein metrics. Writing the metric
\beq
g=4l^2\,\frac uc\,\si_1^2+\eps\,\frac cu\,dt^2+c(\si_2^2+\si_3^2),\eeq
with
\beq
c=t^2-\eps\,l^2,\qq\quad u=mt+\la\left(\eps l^4+2l^2t^2-\eps\frac{t^3}{3}\right),\eeq
and taking  the obvious tetrad, we found the following Killing-Yano tensor
\beq
Y=\eps\,l\,e^0\wedge e^1+t\,e^2\wedge e^3,\eeq
the square of which produces the Killing-St\"ackel tensor
\beq
S=c\Big((e^2)^2+(e^3)^2\Big).\eeq
Taking for hamiltonian
\beq
2H=g^{ij}\,\Pi_i\,\Pi_j=\frac 1{4l^2}\frac{c_{\eps}}{u_{\eps}}\,\Pi_x^2
+\eps\frac{u_{\eps}}{c_{\eps}}\,\Pi_t^2+\frac 1{c_{\eps}}\Big(\Pi_y^2+(\Pi_z-y\Pi_x)^2\Big),
\eeq
the Killing vectors induce observables linear in the momenta
\beq\label{2cK}
\wti{\cal L}_1=\Pi_x,\qq \wti{\cal L}_2=\Pi_y-z\Pi_x,\qq \wti{\cal L}_3=\Pi_z,\qq 
\wti{\cal L}_4=y\,\Pi_z-z\,\Pi_y-\frac 12(y^2-z^2)\Pi_x, 
\eeq
which are conserved
\beq
\{H,\wti{\cal L}_i\}=0,\ i=1,\ldots 4,\eeq
while the Killing-St\"ackel tensor induces a conserved observable which is quadratic in the momenta
\beq
{\cal S}=\Pi_y^2+(\Pi_z-y\Pi_x)^2,\qq\quad\{H,{\cal S}\}=0,\eeq
which is not reducible to a bilinear form with respect to the Killing vectors (\ref{2cK}).

This dynamical system is therefore integrable since $\,H,\,{\cal S},\,\Pi_x,\,\Pi_z$ are in 
involution for the Poisson bracket. Writing the action as
\beq
S=Et+p\,x+q\,z+A(t),\qq p=\Pi_x,\quad q=\Pi_z,\eeq
the Hamilton-Jacobi equation separates and we end up with
\beq
\eps\frac{u}{c}\left(\frac{dA}{dt}\right)^2=
\frac{c}{u}\,\frac{p^2}{4l^2}+\frac{{\cal S}}{c}-2E.\eeq

\section{Type III metrics}
In this case the Lie algebra is defined as
\beq\label{3lie}
[{\cal L}_1,{\cal L}_2]=0,\qq [{\cal L}_2,{\cal L}_3]=0,\qq [{\cal L}_3,{\cal L}_1]={\cal L}_3.\eeq
A representation by differential operators is
\beq\label{3iso}
{\cal L}_1=\pt_x+z\,\pt_z,\qq {\cal L}_2=\pt_y,\qq {\cal L}_3=\pt_z,\eeq
and the invariant Maurer-Cartan 1-forms are
\beq\label{3mc}
\si_1=dx,\quad\si_2=dy,\quad \si_3=e^{-x}\,dz,\quad\Longrightarrow\quad  
d\si_1=d\si_2=0,\quad d\si_3=\si_3\wedge\si_1.\eeq
We will look for diagonal metrics of the form 
\beq\label{3metB}
g=\be^2\,\si_1^2+\ga^2\,\si_2^2+\de^2\,\si_3^2+\eps\alf^2\,dt^2.
\eeq
If $\,\be^2=\de^2$ the metric exhibits a fourth Killing vector
\beq\label{extraK}
{\cal L}_4=z\,\pt_x+\frac 12(z^2-e^{2x})\pt_z,\eeq
and the algebra closes up to
\beq
[{\cal L}_1,{\cal L}_4]={\cal L}_4,\qq [{\cal L}_3,{\cal L}_4]={\cal L}_1.
\eeq

\subsection{Flat space}
For future use let us look for flat space within our 
cooordinates choice. An easy computation shows that it is given by
\beq\label{plateb3}
g_0=\si_2^2+t^2(\si_1^2+\si_3^2)-dt^2=dy^2+t^2\,\frac{(dz^2+dr^2)}{r^2}-dt^2,\qq r=e^x.
\eeq
The flattening coordinates are
\[x_1=y,\qq x_2=\frac{tz}{r},\qq x_3=\frac t{2r}(-1+z^2+r^2),\qq \tau=\frac t{2r}(1+z^2+r^2),\]
which gives
\[g_0=d\vec{r}\cdot d\vec{r}-d\tau^2,\qq\quad \vec{r}=(x_1,x_2,x_3).\]

\subsection{Integration of the field equations}
Following the same procedure as for the type II case, we obtain for independent equations
\[\barr{ll} 
(I)\quad &\dst \frac{\dot{\de}}{\de}=\frac{\dot{\be}}{\be}, \\[5mm]  
(II)\quad & \dst\frac{\ddot{\be}}{\be}+\frac{\dot{\be}}{\be}\left(\frac{\dot{\be}}{\be}
+\frac{\dot{\ga}}{\ga}-\frac{\dot{\alf}}{\alf}\right)+\eps\left(\frac 1{\be^2}+\la\right)\alf^2=0,\\[5mm]
(III)\quad & \dst \frac{\ddot{\ga}}{\ga}+\frac{\dot{\ga}}{\ga}\left(2\frac{\dot{\be}}{\be}
-\frac{\dot{\alf}}{\alf}\right)+\eps\,\la\,\alf^2=0, \\[5mm]    
(IV)\quad & \dst \frac{\dot{\be}^2}{\be^2}+2\frac{\dot{\be}\dot{\ga}}{\be\ga}
+\eps\left(\frac 1{\be^2}+\la\right)\alf^2=0.\earr\]
Relations (I) and (II)-(IV) integrate up to \footnote{The coefficient between $\be$ and $\de$ can be set to 1 by rescaling the coordinate $z$. }
\beq\label{int1}
\dot{\be}=c\,\alf\,\ga,\qq c\in{\mb R},\qq\qq \de=\be.\eeq
The time coordinate choice 
\[\alf=\frac{\be}{\ga}\quad\Longrightarrow\quad \de=\be=\be_0\,e^{c t}.\]
To determine $\ga$ we have to use (III) which becomes 
\beq\label{int2}
\frac{\ddot{\ga}}{\ga}+\frac{\dot{\ga}^2}{\ga^2}+c\,\frac{\dot{\ga}}{\ga}+
\eps\la\be_0^2\,\frac{e^{2c t}}{\ga^2}=0.
\eeq
This equation does linearize in $\ga^2$ to
\beq\label{int3}
\ddot{(\ga^2)}+c\,\dot{(\ga^2)}+2\eps\la\be_0^2\,e^{2c t}=0,
\eeq
and the remaining relation (IV) becomes
\beq\label{int4}
c\,\dot{(\ga^2)}+c^2\,\ga^2+\eps(1+\la\,\be^2)=0.
\eeq
Let us organize the discussion according to the values of $c$.

\subsection{The metrics}
We will consider first the special case $c=0$. Relation (\ref{int4}) gives 
$\,\be^2=-1/\la$ and (\ref{int3}) is easily integrated to $\ga^2=\ga_0+\ga_1 t+\eps t^2$. 
By a translation of $t$ we can set $\ga_1\to 0$ and by a rescaling of $z$ we can set $c_2\to 1$, 
so we can write the metric
\beq\label{MEfin1} 
g=\frac 1{|\la|}\Big[\si_1^2+\si_3^2+\ga^2\,\si_2^2+\eps\frac{dt^2}{\ga^2}\Big],\qq 
\ga^2=\ga_0+\eps t^2,\qq \la<0.
\eeq
Let us emphasis that all the metrics will have negative Einstein constant.

For the minkowskian signature we must have $\ga_0>0$. By a scaling of the variables $y$ and $t$ we can 
set $\ga_0=1$. This leaves us with 
\beq
g=\frac 1{|\la|}\left\{dx^2+e^{-2x}\,dz^2+(1-t^2)\,dy^2-\frac{dt^2}{(1-t^2)}\right\}.
\eeq
The change of coordinates
\beq
\mu=\frac 12\Big[e^x+(1+z^2)\,e^{-x}\Big],\qq \tan\phi=\frac 1{2z}\Big[e^{2x}-(1-z^2)\Big],
\eeq 
leads to
\beq
g=\underbrace{\frac 1{|\la|}\left\{\frac{d\mu^2}{\mu^2-1}+(\mu^2-1)\,d\phi^2\right\}}_{g_0}+
\underbrace{\frac 1{|\la|}\left\{(1-t^2)\,dy^2-\frac{dt^2}{1-t^2}\right\}}_{g_1},\eeq
on which we recognize a product of 2-dimensional Einstein metrics with the {\em same} scalar curvature: 
the euclidean $g_0={\mb H}^2$ and the lorentzian  $g_1=AdS_2$, so we end up with 6 Killing vectors. 
Let us notice that this it is a well known fact \cite{Be}[p. 44] that for a product to be Einstein, it is mandatory that both two dimensional metrics in the product have the same Einstein constant.

For the euclidean signature, according to the sign of $\ga_0$ we have 3 cases:
\[\barr{ll}
\ga_0>0\qq & \dst g=\frac 1{|\la|}\left\{\si_1^2+\si_3^2
+\frac 1{\cos^2 \tau}\Big[\si_2^2+d\tau^2\Big]\right\},\\[5mm]
\ga_0=0\qq & \dst g=\frac 1{|\la|}\left\{\si_1^2+\si_3^2+\frac 1{\tau^2}\Big[\si_2^2+d\tau^2\Big]\right\},\\[5mm] 
\ga_0<0\qq & \dst  g=\frac 1{|\la|}\left\{\si_1^2+\si_3^2
+\frac 1{{\sinh}^2\,\tau}\Big[\si_2^2+d\tau^2\Big]\right\},
\earr\qq\quad \la<0.\]
We have again decomposable Einstein metrics made up of two copies of ${\mb H}^2$.

Let us consider the more general case for which $c$ does not vanish. We obtain  
\beq
\tilde{f}(t)\equiv c_2^2\,\ga^2=-\eps+\ga_1\,e^{-c t}-\frac{\eps\la\be_0^2}{3}\,e^{2c t}.
\eeq
Taking as variable $s=\be_0\,e^{c t}$, and cleaning up the irrelevant parameters, 
we eventually obtain the Einstein metric
\beq\label{MEfin2}
g=s^2(\si_1^2+\si_3^2)+f(s)\,\si_2^2+\eps\,\frac{ds^2}{f(s)},\qq
f(s)=-\eps+\frac{\ga_0}{s}-\frac{\eps\la}{3}\,s^2.
\eeq
The metric exhibits the extra Killing vector (\ref{extraK}) but is no longer decomposable. It 
contains two essential constants: $\ga_0$ and $\la$ which are expected for the general solution.

For $\eps=-1$ it was first obtained by Stewart and Ellis \cite{se} and 
by Cahen and Defrise \cite{cd} and re-discovered later on in \cite{mt}, \cite{mm} and 
more recently in \cite{ctbis}. In \cite{sk} the metric is given  by formulas (13.9)and (13.48) 
in which one has to take $e=l=0$ and $k=-1$. The Weyl tensor has a single non-vanishing 
component $\dst\Psi_2=-\frac c{2s^3}$ giving Petrov type D.

For $\eps=+1$, this metric was obtained by Lorenz-Petzold \cite{Lo}.  Using the obvious vierbein, 
we obtain for the Weyl tensor 
\beq\label{3Weyl}
W^+=W^-=\frac{\ga_0}{2s^3}\,M,\qq\quad M={\rm diag}(1,-2,1),\eeq
giving Petrov type $(D^+,\,D^-)$.The Weyl tensor is self-dual if and only if $\ga_0=0$.

In this case, for negative $\la$, we obtain a complete Einstein metric in the following way: let us 
change the variable $s$ into $u=\sqrt{|\la|}\,s$. Then 
\[f(s)\to h(u)=\frac 1{3u}\,(u^3-3u+2c),\qq\quad 2c=3\sqrt{|\la|}\,\ga_0,\]
so that the choice $c=-1$ gives a double root and for metric :
\beq
g=\frac 1{|\la|}\left(u^2(\si_1^2+\si_3^2)+|\la|\,h\,dy^2+\frac{du^2}{h}\right),
\qq h(u)=\frac{(u-2)}{3u}(u+1)^2.\eeq
Positivity requires $u>2$ and the metric becomes singular for $u=2$. That this singularity 
is only apparent follows from a local analysis. If we take as new variables:
\[\xi\approx \sqrt{\frac 83(u-2)}\to 0,\qq \tilde{y}=\frac 34\sqrt{|\la|}\,y,\]
the local form of the metric becomes a product metric $\,{\mb H}^2\times {\mb R}^2$
\[g\approx \frac 1{|\la|}\Big(4(\si_1^2+\si_3^2)+\xi^2\,d\tilde{y}^2+d\xi^2\Big),
\qq \tilde{y}\in\,[0,2\pi].\]
The $u=2$ singularity is therefore a removable polar-like singularity.  

\subsection{The special case $\ga_0=0$}
From (\ref{3Weyl}) we see that for $\ga_0=0$ the metric is conformally flat, so we must recover 
symmetric spaces with 10 Killing vectors instead of 4. 

Let us first consider the minkowskian signature for $\la>0$. We can write the metric
\beq\label{dSB3}
g^+_M=\frac 3{\la}\left[t^2(dx^2+e^{-2x}\,dz^2)-\frac{dt^2}{1+t^2}+(1+t^2)\,du^2\right],\qq
t=\sqrt{\frac{\la}{3}}\,s,\quad u=\sqrt{\frac{\la}{3}}\,y.\eeq
The coordinates  
\beq\left\{\barr{l}
z^1=tz\,e^{-x},\quad z^2=t(\sinh x+e^{-x}\,z^2/2),\quad z^3=\sqrt{1+t^2}\,\cos u,\\[4mm]  
z^0=t(\cosh x+e^{-x}\,z^2/2),\quad z^4=\sqrt{1+t^2}\,\sin u,\earr\right.
\eeq
are constrained by $\,(z^1)^2+(z^2)^2+(z^3)^2-(z^0)^2+(z^4)^2=1$ and
\beq
g^+_M=\frac 3{\la}\Big((dz^1)^2+(dz^2)^2+(dz^3)^2-(dz^0)^2+(dz^4)^2\Big),\eeq
which is de Sitter metric, with isometry group enhanced to $O(4,1)$.

For $\la<0$ we start from
\beq
g^-_M=\frac 3{|\la|}\left[t^2(dx^2+e^{-2x}\,dz^2)-\frac{dt^2}{1-t^2}+(1-t^2)\,du^2\right],
\quad t=\sqrt{\frac{|\la|}{3}}\,s,\quad u=\sqrt{\frac{|\la|}{3}}\,y.\eeq
The coordinates 
\beq\left\{\barr{l}
z^1=tz\,e^{-x},\quad z^2=t(\sinh x+e^{-x}\,z^2/2),\quad z^3=\sqrt{1+t^2}\,\cosh u,\\[4mm]  
z^0=t(\cosh x+e^{-x}\,z^2/2),\quad z^4=\sqrt{1+t^2}\,\sinh u,\earr\right.
\eeq
are constrained by $\,(z^1)^2+(z^2)^2-(z^3)^2-(z^0)^2+(z^4)^2=1$ and
\beq
g^-_M=\frac 3{\la}\Big((dz^1)^2+(dz^2)^2-(dz^3)^2-(dz^0)^2+(dz^4)^2\Big),\eeq
and the isometry group is enhanced to $O(3,2)$.

For the euclidean signature, positivity requires $\la<0$. We start from
\beq\label{dSb3}
g^-_E=\frac 3{|\la|}\left[t^2(dx^2+e^{-2x}\,dz^2)-\frac{dt^2}{t^2-1}+(t^2-1)\,du^2\right],
\quad t=\sqrt{\frac{|\la|}{3}}\,s,\quad u=\sqrt{\frac{|\la|}{3}}\,y.\eeq
The coordinates 
\beq\left\{\barr{l}
z^1=tz\,e^{-x},\quad z^2=t(\sinh x+e^{-x}\,z^2/2),\quad z^3=\sqrt{t^2-1}\,\cos u,\\[4mm]  
z^0=t(\cosh x+e^{-x}\,z^2/2),\quad z^4=\sqrt{t^2-1}\,\sin u,\earr\right.
\eeq
are constrained by $\,(z^1)^2+(z^2)^2+(z^3)^2-(z^0)^2+(z^4)^2=-1$ and
\beq
g^-_M=\frac 3{\la}\Big((dz^1)^2+(dz^2)^2-(dz^3)^2-(dz^0)^2+(dz^4)^2\Big),\eeq
and the metric lies on the manifold ${\mb H}^4$.

The de Sitter metric (\ref{dSb3}) is written in quite ``exotic" coordinates. Since this 
metric is of some importance, we give in appendix A, the explicit form of its Killing 
vectors.

\subsection{Integrable geodesic flow}
With the obvious tetrad, we found the following Killing-Yano and Killing-Stackel tensors
\beq
Y=s\,e^3\wedge e^1,\qq\Rightarrow\qq S=s^2\Big((e^1)^2+(e^3)^2\Big).\eeq
Let us consider the geodesic flow induced by the Hamiltonian 
\beq
2H\equiv g^{ij}\,\Pi_i\,\Pi_j=\frac 1{f(s)}\,\Pi_y^2+\frac{\Pi_x^2+e^{2x}\Pi_z^2}{s^2}
+\eps\,f(s)\,\Pi_s^2.
\eeq
The KS tensor $\,S$ gives for conserved quantity
\beq
{\cal S}=\Pi_x^2+e^{2x}\Pi_z^2,\qq\quad \{H,{\cal S}\}=0.
\eeq
It cannot be obtained from symmetrized tensor products of Killing vectors because their 
corresponding linear conserved quantities are
\[\wti{\cal L}_1=\Pi_x+z\,\Pi_z, \quad \wti{\cal L}_2=\Pi_y,\quad \wti{\cal L}_3=\Pi_z,\quad 
\wti{\cal L}_4=z\,\Pi_x+\frac 12\Big(z^2-e^{2x}\Big)\Pi_z,\quad \{H,\wti{\cal L}_i\}=0.\]
The dynamical system with hamiltonian $H$ is therefore integrable, since it exhibits 4 
independent conserved quantities: $\,H,\,{\cal S},\,\Pi_y,\,\Pi_z$ in involution for the 
Poisson bracket. Writing the action as
\beq
S=E\,t+p\,y+q\,z+A(s),\qq\quad p=\Pi_y,\quad q=\Pi_z
\eeq 
we get for separated Hamilton-Jacobi equation
\beq
\left(\frac{dA}{ds}\right)^2=\eps\left(\frac{2E}{f}-\frac{\cal S}{s^2\,f}-\frac{p^2}{f^2}\right).
\eeq

Let us consider now the Bianchi V case.

\section{Type V metrics}
In this case the Lie algebra is
\beq\label{5lie}
[{\cal L}_1,{\cal L}_2]={\cal L}_2,\qq [{\cal L}_2,{\cal L}_3]=0,\qq [{\cal L}_3,{\cal L}_1]=-{\cal L}_3,\eeq
with the Killing vectors 
\beq\label{5iso}
{\cal L}_1=\pt_x-y\pt_y-z\pt_z,\qq {\cal L}_2=\pt_y,\qq {\cal L}_3=\pt_z,\eeq
and the invariant Maurer-Cartan 1-forms  
\beq\label{5mc}
\si_1=dx,\ \si_2=e^x\,dy,\  \si_3=e^x\,dz,\quad\Rightarrow\quad 
d\si_1=0,\quad d\si_2=\si_1\wedge\si_2,\quad d\si_3=\si_1\wedge\si_3.\eeq
We will look again for a diagonal metric of the form (\ref{3metB}).

\subsection{The flat space}
Let us first determine the flat space Bianchi V metric. It is easy to check that it is given by
\beq\label{b5plate}
g_0=t^2(\si_1^2+\si_2^2+\si_3^2)-dt^2=t^2\,\ga-dt^2,
\eeq
where the metric $\ga$ is the Poincar\'e metric for ${\mb H}^3$:
\[\ga\equiv \si_1^2+\si_2^2+\si_3^2=\frac{dy^2+dz^2+d\rho^2}{\rho^2},\qq \rho=e^{-x},\]
which has 6 Killing vectors. The flattening coordinates 
for (\ref{b5plate}) are
\beq\label{b5plate1}
x_1=\frac{ty}{\rho},\quad x_2=\frac{tz}{\rho},\quad x_3=\frac t{2\rho}(-1+y^2+z^2+\rho^2),
\quad \tau=\frac t{2\rho}(1+y^2+z^2+\rho^2),\eeq
leading to 
\beq\label{b5plate2}
g_0\equiv t^2(\si_1^2+\si_2^2+\si_3^2)-dt^2=d\vec{r}\cdot d\vec{r}-d\tau^2,
\qq \vec{r}=(x_1,x_2,x_3).\eeq

\subsection{Integration of the field equations}
The independent equations are now
\beq\label{b5E}\barr{ll}
(I)\quad & \dst\frac{\ddot{\be}}{\be}+\frac{\dot{\be}}{\be}
\left(\frac{\dot{\ga}}{\ga}+\frac{\dot{\de}}{\de}-\frac{\dot{\alf}}{\alf}\right)
+\eps(2+\la\,\be^2)\frac{\alf^2}{\be^2}=0,\\[5mm]
(II)\quad & \dst\frac{\ddot{\ga}}{\ga}+\frac{\dot{\ga}}{\ga}
\left(\frac{\dot{\be}}{\be}+\frac{\dot{\de}}{\de}-\frac{\dot{\alf}}{\alf}\right)
+\eps(2+\la\,\be^2)\frac{\alf^2}{\be^2}=0,\\[5mm]
(III)\quad & \dst\frac{\ddot{\de}}{\de}+\frac{\dot{\de}}{\de}
\left(\frac{\dot{\be}}{\be}+\frac{\dot{\ga}}{\ga}-\frac{\dot{\alf}}{\alf}\right)
+\eps(2+\la\,\be^2)\frac{\alf^2}{\be^2}=0,\\[5mm]
(IV)\quad & \dst \frac{\dot{\be}\dot{\ga}}{\be\ga}+\frac{\dot{\ga}\dot{\de}}{\ga\de}+
\frac{\dot{\be}\dot{\de}}{\be\de}+\eps(3+\la\,\be^2)\frac{\alf^2}{\be^2}=0,\earr  \quad (V)\quad 
\frac{\dot{\de}}{\de}-2\frac{\dot{\be}}{\be}+\frac{\dot{\ga}}{\ga}=0.
\eeq
The differences (I)-(II) and (I)-(III) integrate to
\[\frac{\dot{\ga}}{\ga}-\frac{\dot{\be}}{\be}=c\,\frac{\alf}{\be\ga\de},\qq\quad 
\frac{\dot{\de}}{\de}-\frac{\dot{\be}}{\be}=c_2\,\frac{\alf}{\be\ga\de},\]
and (V) implies $c_2=-c$.

This suggests to fix up the time coordinate by imposing 
\[
\alf=\be\ga\de\quad\Longrightarrow\quad \ga=\ga_0\,e^{c t}\,\be,\qq \de=\de_0\,e^{-c t}\,\be,\qq \alf=\ga_0\,\de_0\,\be^3.\]
By a rescaling of the coordinates $y$ and $z$, we may set $\ga_0=\de_0=1$ 
and relation (I) becomes
\[
D_t{\left(\frac{\dot{\be}}{\be}\right)}+\eps\,\be^4(2+\la\,\be^2)=0,
\quad\Longrightarrow\quad \frac{\dot{\be}^2}{\be^2}+\eps\,\be^4(1+\la\,\be^2/3)=E.\]
Eventually relation (IV) gives $E=c^2/3\geq 0$. Summarizing, we have obtained for the Einstein metric
\beq\label{5intfin}
g=\be^2\Big(\si_1^2+e^{2c t}\,\si_2^2
+e^{-2c t}\,\si_3^2+\eps\,\,\be^4\,dt^2\Big),\qq 
\frac{\dot{\be}^2}{\be^2}=\frac{c^2}{3}-\eps\,\be^4(1+\la\,\be^2/3).\eeq
For the minkowskian signature, this result was first obtained by Sch\"ucking and Heckmann 
\cite{sh} and written, in \cite{sk}[p. 192] as
\[g=-d\tau^2+S^2(\tau)\Big(\si_1^2+F^{\sqrt{3}}\,\si_2^2+F^{-\sqrt{3}}\,\si_3^2\Big),\]
with the relations
\[
3\left(\frac{dS}{d\tau}\right)^2=3+\frac{\Sigma^2}{S^4}+\la\,S^2,\qq\quad
F^{\sqrt{3}}=\exp\left(2\Sigma\int\frac{d\tau}{S^3(\tau)}\right).\]
Upon the identifications 
\[d\tau=\be^3(t)\,dt,\qq S(\tau)=\be(t),\qq\Sigma=c,\]
the differential equation for $S(\tau)$ gives the differential equation for $\be(t)$  
and for $F$ we get $\dst F^{\sqrt{3}}=e^{2ct}$ showing full agreement with (\ref{5intfin}).

\subsection{The special case $\,E=c=0$}
This special case leads to metrics with enhanced symmetries, namely non-compact symmetric spaces with 
10 Killing vectors. Among these, as mentioned in \cite{sk}, we expect de Sitter metrics. 

The differential equation (\ref{5intfin}) becomes
\beq
dt=\frac{d\be}{\be^3\sqrt{-\eps-\eps\la\be^2/3}}.\eeq
Taking $\be\to s$ as a  new variable, we get the metric
\beq\label{Esimple}
g=s^2\Big(\si_1^2+\si_2^2+\si_3^2\Big)-\frac{ds^2}{1+\frac{\dst\la\,s^2}{3}}.
\eeq
The minkowskian or euclidean character of the metric does depend solely on the range taken by the 
variable $t$, and in the $\la\to 0$ limit we recover, as it should, the flat space metric (\ref{b5plate}). 

For $\la>0$, we can have only a minkowskian metric. As explained at the beginning of this section, 
we expect a higher symmetry. Defining $\dst \sqrt{\frac{\la}{3}}\,s=\frac{2t}{1-t^2}$ 
we can write the metric:
\[g^+_M=\frac{12}{\la}\frac 1{(1-t^2)^2}\Big(t^2(\si_1^2+\si_2^2+\si_3^2)-dt^2\Big),\]
on which we recognize a symmetric space, since by using the flattening coordinates (\ref{b5plate1}), we have
\[g^+_M=\frac{12}{\la}\frac{d\vec{r}\cdot d\vec{r}-d\tau^2}{(1+\vec{r}\,^2-\tau^2)^2}.\]
Indeed, using the constrained coordinates 
\[z_0=\frac{1-\vec{r}\,^2+\tau^2}{1+\vec{r}\,^2-\tau^2},\qq 
\vec{z}=\frac{2\vec{r}}{1+\vec{r}\,^2-\tau^2},
\qq z_4=\frac{2\tau}{1+\vec{r}\,^2-\tau^2},\qq z_0^2+\vec{z}\,^2-z_4^2=1,\]
we see that we end up with de Sitter metric
\[g^+_M=\frac 3{\la}\Big(dz_0^2+d\vec{z}\cdot d\vec{z}-dz_4^2\Big),\]
and the isometry group enlarges to $O(4,1)$. As for Bianchi type III, we get de Sitter metric 
in some ``exotic" coordinates. Since this could be perhaps useful for other 
applications, we give in Appendix A the form of its Killing vectors.

For $\la<0$ we have, for Minkowskian signature, anti de Sitter metric
\[ g^-_M=\frac{12}{|\la|(1+t^2)^2}\,\Big(t^2(\si_1^2+\si_2^2+\si_3^2)-dt^2\Big).\]
For the euclidean signature we get
\[g^-_E=\frac 3{|\la|}\Big[{\rm ch}\,^2\tht(\si_1^2+\si_2^2+\si_3^2)+d\tht^2\Big],\]
which is also a locally symmetric space. To give the embedding in ${\mb R}^5$ let us first 
define a set of 4 coordinates $(\vec{r},\,x^0)$ by
\[\vec{r}=\Big(e^x\,y,\ e^x\,z,\ -\sinh x+e^x(y^2+z^2)/2\Big),\qq x^0=\cosh x+e^x(y^2+z^2)/2,\]
which are constrained by $\,\vec{r}\,^2-(x^0)^2=-1$. One can check that
\[\si_1^2+\si_2^2+\si_3^2=dx^2+e^{2x}(dy^2+dz^2)=d\vec{r}\cdot d\vec{r}-(dx^0)^2.\]
Then, defining the coordinates $\,(\vec{z},\,z^0,\,z^4)$ by 
\[\vec{z}=\cosh\tht\,\vec{r},\qq z^0=\cosh\tht,\qq z^4=\sinh\tht,\]
we conclude to
\beq
g^-_E=\frac 3{|\la|}\Big(d\vec{z}\cdot d\vec{z}-(dz^0)^2+(dz^5)^2\Big),\qq \vec{z}\,^2-(z^0)^2+(z^4)^2=-1,
\eeq
which shows that the metric $g^-_E$ lives on the manifold ${\mb H}^4$.

\subsection{The general case $\,E\neq 0$}
In relation (\ref{5intfin}), let us introduce as a new variable
\beq\label{deft}
\rho=\frac{|c|}{\be^2}>0\quad\Longrightarrow\quad  
\frac{\rho\,d\rho}{\sqrt{P(\rho)}}=\pm\frac{2c\,dt}{\sqrt{3}},\qq\qq P(\rho)\equiv \rho(\rho^3-3\eps\,\rho-\eps \la |c|),
\eeq
which gives for the metric
\beq\label{finmet}
g=\frac{|c|}{\rho}\left(\si_1^2+\ga^2\,\si_2^2+\frac 1{\ga^2}\,\si_3^2
+\frac 34\,\eps\,\frac{d\rho^2}{P(\rho)}\right),\qq\quad \ga^2\equiv e^{2|c|t}.
\eeq

\nin{\bf Remark:} Due to the symmetric role played by $\,(\si_2,\,\si_3)$, the coefficients of 
$\si_2^2$ and of $\si_3^2$ may be interchanged and this corresponds to the exchange 
$\,(c\leftrightarrow -c)$ or $\,\dst\Big(\ga\leftrightarrow\frac 1{\ga}\Big)$. This means that if 
the metric (\ref{finmet}) is Einstein, then 
\beq
g=\frac{|c|}{\rho}\left(\si_1^2+\frac 1{\ga^2}\,\si_2^2+\ga^2\,\si_3^2
+\frac 34\,\eps\,\frac{d\rho^2}{P(\rho)}\right),
\eeq
will be Einstein too. We will use this observation to get rid of the sign in relation (\ref{deft}) and to take $c>0$.

\subsection{Minkowskian signature}
In \cite{sh} the results are given up to the quadrature for $\be(t)$. For the Einstein metric 
of interest this quadrature requires the use of elliptic functions. The technical 
details are given in the appendix; using these results we get the final form of the metrics, 
according to the sign of the Einstein constant. 

We have to take $\eps=-1$. In this case the cubic polynomial $P(\rho)=\rho(\rho^3+3\rho+\la c)$ has, 
no matter what the value of $c$ is, always 2 real and 2 complex conjugate roots (recall that we 
exclude $\la=0$). So we fix $c=1$ and, to express most conveniently the roots of $P$, we 
parametrize the Einstein constant according to
\[\la=2\,\sinh(\tht),\qq\qq \tht\in\,{\mb R}\backslash \{0\}.\]
We will use now the results from appendix B to give the explicit form of the metric.
\brm
\item \fbox{For $\la<0$ :} 

\nin In this case the roots are
\[a=-2\,\sinh(\tht/3)\ >\ b=0,\qq a_1=\sqrt{3}\,\cosh(\tht/3),\quad  
b_1=\sinh(\tht/3),\]
so we have
\[\left\{\barr{l}\dst 
A=\sqrt{3+12\,\sinh^2(\tht/3)},\\[4mm]\dst  B=\sqrt{3+4\,\sinh^2(\tht/3)},\earr\right.  
\qq\quad k^2=\frac{(A+B)^2-4\,\sinh^2(\tht/3)}{4AB}\]
and
\[{\rm sn}\,v_0=\sqrt{\frac{2B}{A+B-2\,\sinh(\tht/3)}}.\]
In formula (\ref{finmet}) we have to transform $d\rho$ into $dv$ to get eventually
\beq\label{finmet2}
g_M=\frac 1{\rho}\left(\si_1^2+\ga^2\,\si_2^2+\frac 1{\ga^2}\,\si_3^2
-\frac 3{AB}\,(dv)^2\right),\qq\quad v\in\,[0,v_0),
\eeq
where $\rho$ and $\ga^2$ are given respectively by
\beq
\rho=\frac{aB\,{\rm cn}^2\,v}{B\,{\rm cn}^2\,v-A\,{\rm sn}^2\,v\,{\rm dn}^2\,v},
\eeq
and by
\beq
\ga^2=\left(e^{-\xi v}\frac{H(v_0+v)\,\T_1(v_0+v)}{H(v_0-v)\,\T_1(v_0-v)}\right)^{\sqrt{3}},\qq 
\xi=2\left(\frac{\T'}{\T}(v_0)+\frac{H_1'}{H_1}(v_0)\right).
\eeq

\item \fbox{For $\la>0$ :} 

\nin In this case the roots are
\[a=0\ >\ b=-2\,\sinh(\tht/3),\qq a_1=\sqrt{3}\,\cosh(\tht/3),\quad  
b_1=\sinh(\tht/3),\]
so we have
\[\left\{\barr{l}\dst 
A=\sqrt{3+4\,\sinh^2(\tht/3)},\\[4mm]\dst  B=\sqrt{3+12\,\sinh^2(\tht/3)},\earr\right.  
\qq\quad k^2=\frac{(A+B)^2-4\,\sinh^2(\tht/3)}{4AB}.\]
The parameter $k^2$ remains unchanged while $A$ and $B$ are interchanged and $v_0$ becomes
\[{\rm sn}\,v_0=\sqrt{\frac{2B}{A+B+2\,\sinh(\tht/3)}}.\]
The metric is still given by (\ref{finmet2}), where now $\rho$ and $\ga^2$ are respectively 
\beq
\rho=\frac{|b|A\,{\rm sn}^2\,v\,{\rm dn}^2\,v}{B\,{\rm cn}^2\,v-A\,{\rm sn}^2\,v\,{\rm dn}^2\,v},
\eeq
and by
\beq
\ga^2=\left(e^{-\xi v}\frac{H(v_0+v)\,\T_1(v_0+v)}{H(v_0-v)\,\T_1(v_0-v)}\right)^{\sqrt{3}},\qq 
\xi=2\left(\frac{|b|}{AB}+\frac{\T'}{\T}(v_0)+\frac{H_1'}{H_1}(v_0)\right).\eeq
\erm
Using the complex null-tetrad \footnote{We follow strictly the notations of \cite{sk}.}
\beq
m=\frac 1{\sqrt{2}}(e^{ct}\,\be\,\si_2+ie^{-ct}\,\be\,\si_3),\qq k=\frac 1{\sqrt{2}}(\be^3\,dt-\be\,\si_1),\qq 
l=\frac 1{\sqrt{2}}(\be^3\,dt+\be\,\si_1),\eeq
and defining $\mu=1/\be^2$, one has to use the differential equation (\ref{5intfin}) which gives
\beq
\frac{\dot{\mu}^2}{4}=\frac{c^2}{3}\,\mu^2+1+\frac{\la}{3}\,\frac 1{\mu}.\eeq
Just using these informations one can check, computing the curvature, the Einstein property of 
this metric. For the Weyl tensor we have obtained
\beq
\Psi_0=c\,\mu^2\Big(1-\frac{\dot{\mu}}{2}\Big),\quad \Psi_2=\frac{c^2}{3}\,\mu^3,\quad \Psi_4=-c\,\mu^2\Big(1+\frac{\dot{\mu}}{2}\Big),\qq 
\Psi_1=\Psi_3=0,\eeq
which establishes the Petrov type I of the metric.

\subsection{Euclidean signature}
In this case $\,P(\rho)=\rho(\rho^3-3\rho-\la c)$.  It has two real roots for 
$\,\la c\in(-\nf,-2)\cup (+2,+\nf)$, four real roots for $\la c\in[-2,0)\cup(0,+2]$ and 
a double root for $\la c=\pm 2$. Since the parameter $c$ is free, we can collapse 
$\,(-\nf,0)\cup (0,+\nf)$ to two points by taking $c=2/|\la|$. Therefore in this case elliptic functions 
are no longer required!  

We have to discuss two cases:
\brm
\item \fbox{$\la<0$ :}

\nin We have $P(\rho)=\rho(\rho+2)(\rho-1)^2$ and
\[\frac{2c}{\sqrt{3}}\ dt=\frac{\rho\,d\rho}{|\rho-1|\sqrt{\rho(\rho+2)}}.\]
The change of variable $\dst\rho=\frac{2s^2}{3-s^2}$ simplifies to
\[2c\ dt=\frac{4s^2\,ds}{(1-s^2)(3-s^2)}.\]
We obtain
\beq
\ga^2\equiv e^{2ct}=\frac{1+s}{|1-s|}\left(\frac{\sqrt{3}-s}{\sqrt{3}+s}\right)^{\sqrt{3}},
\eeq
and the Einstein metric  
\beq
g_E=\frac{(3-s^2)}{|\la|\,s^2}\left(\si_1^2+\ga^2\,\si_2^2+\frac 1{\ga^2}\,\si_3^2
+\frac{ds^2}{(1-s^2)^2}\right).
\eeq
In fact we have two different metrics, according to the interval taken for $s$: either  $\,s\in(-1,1)$ or $\,s\in(1,\sqrt{3})$.

\item \fbox{$\la>0$ :}

\nin We have $P(\rho)=\rho(\rho-2)(\rho+1)^2$ and
\[\frac{2c}{\sqrt{3}}\ dt=\frac{\rho\,d\rho}{(\rho+1)\sqrt{\rho(\rho+2)}},\qq\quad \rho>2.\]
The change of variable $\dst\rho=\frac 2{1-s^2}$ simplifies to
\[\frac{2c}{\sqrt{3}}\ dt=-\frac{4\,ds}{(1-s^2)(3-s^2)},\qq s\in(-1,+1).\]
Deleting the sign we obtain
\beq
\ga^2\equiv e^{2c\,t}=\frac{\sqrt{3}-s}{\sqrt{3}+s}\left(\frac{1+s}{1-s}\right)^{\sqrt{3}}
\eeq
and the Einstein metric 
\beq
g_E=\frac{(1-s^2)}{\la}\left[\si_1^2+\ga^2\,\si_2^2+\frac 1{\ga^2}\,\si_3^2
+\frac{3\,ds^2}{(3-s^2)^2}\right]. 
\eeq
\erm
Some remarks are now in order:
\brm
\item We have checked, using the obvious vierbein, the vanishing of the matrix $\,B$ and that ${\rm Tr}\,A=\la$, which proves the Einstein character of both metrics and computed the Weyl tensor: it has Petrov type $\,(I^+,I^-)$ and is never self-dual.
\item It is interesting to compare with the results in \cite{Tobis} for the Bianchi type A Einstein metrics with self-dual Weyl tensor: except for Bianchi type II, they all involve Painlev\'e transcendents.
\item Let us observe that the difference in structure between the minkowskian and euclidean type V case is quite unusual. Indeed we have seen for the Bianchi type II and III metrics that the change 
in the signature brings rather small variance.
\erm

\section{Conclusion}
We have been giving very simple derivations of the ``diagonal" Bianchi type II, III and V Einstein metrics. The first two  exhibit an integrable geodesic flow, while the third one gives rise to new euclidean metrics which can be expressed in terms of elementary functions. Let us observe 
that there is little room for diagonal type B euclidean metrics with self-dual Weyl tensor: we 
have seen that the corresponding metrics are conformally flat, in agreement with \cite{To}. A question of interest is to what extent one could work out the tri-axial type II metric or the 
more general Bianchi VI$_h$ and Bianchi VII$_h$ metrics, a rather difficult aim to say nothing 
of the non-diagonal ones! 

\vspace{2mm}
\nin{\bf Acknowledgements:} we are greatly indebted to Dr Lorenz-Petzold and Pr MacCallum 
for having provided me with the references prior to this work.

\vspace{10mm}
\centerline{\bf\Large Appendix}

\appendix

\section{De Sitter metric re-visited}
\subsection{de Sitter from Bianchi type III}
Let us consider the de Sitter metric (\ref{dSB3}) written as
\beq
g^+_M=\frac 3{\la}\left[t^2\,\frac{dz^2+dv^2}{v^2}+(1+t^2)du^2-\frac{dt^2}{1+t^2}\right],
\qq u=\sqrt{\frac{\la}{3}}\,y,\qq v=e^x.\eeq 
The four standard Killing vectors are now
\beq
K_1=v\,\pt_v+z\,\pt_z,\qq K_2=\pt_u,\qq K_3=\pt_z,\qq K_4=zv\,\pt_v+\frac{(z^2-v^2)}{2}\,\pt_z.\eeq
The remaining ones appear by pairs
\[-\frac{\sin u}{vf}\,\pt_u+\frac{f\,\cos u}{v}(v\,\pt_v+t\,\pt_t),\qq
\frac{\cos u}{vf}\,\pt_u+\frac{f\,\sin u}{v}(v\,\pt_v+t\,\pt_t),\]
\[-\frac{z\sin u}{vf}\,\pt_u+\frac{f\cos u}{v}(zv\,\pt_v-v^2\,\pt_z+zt\,\pt_t),\qq  
\frac{z\cos u}{vf}\,\pt_u+\frac{f\sin u}{v}(zv\,\pt_v-v^2\,\pt_z+zt\,\pt_t),\]
and
\[\left\{\barr{l}\dst
\frac{(v^2+z^2)\sin u}{vf}\,\pt_u
+\frac{f\cos u}{v}\Big((v^2-z^2)v\,\pt_v+2v^2z\,\pt_z-(v^2+z^2)t\,\pt_t\Big),\\[5mm]\dst
-\frac{(v^2+z^2)\cos u}{vf}\,\pt_u
+\frac{f\sin u}{v}\Big((v^2-z^2)v\,\pt_v+2v^2z\,\pt_z-(v^2+z^2)t\,\pt_t\Big),\earr\right.
\  f=\frac{\sqrt{1+t^2}}{t}.\]
Despite the simple form of the metric in these coordinates, the symmetries are somewhat wild.

\subsection{de Sitter from Bianchi type V}
We have shown that the metric
\beq\label{dS1}
g=s^2\Big(\si_1^2+\si_2^2+\si_3^2\Big)-\frac{ds^2}{1+\frac{\dst\la\,s^2}{3}},\qq \la>0,
\eeq
is de Sitter. Taking 
$\sinh\tht=\sqrt{\frac{\la}{3}}\,s$ and $v=e^{-x}$ as new variables the metric becomes
\beq\label{dS2}
g=\frac 3{\la}\Big(\sinh^2 \tht\,\frac{dy^2+dz^2+dv^2}{v^2}-d\tht^2\Big),\qq\quad \la>0.
\eeq
The first 3 dimensional piece in the metric is ${\mb H}^3$ in Poincar\'e coordinates, so we have 
2 sub-algebras:
\beq
{\cal A}_1=\Big\{P_1,\,P_2,\,M_3\Big\},\qq\qq {\cal A}_2=\Big\{Q_1,\,Q_2,\,L_3\Big\}.\eeq
The first one is $e(2)$ ($M_3$ is a rotation) 
\beq
P_1=\pt_y,\quad P_2=\pt_z,\quad M_3=-z\,\pt_y+y\,\pt_z,
\eeq
and the second one is $\tilde{e}(2)$ ($L_3$ is a dilatation)
\beq\left\{\barr{l}
Q_1=\frac 12\Big(-y^2+z^2+v^2\Big)\pt_y-yz\,\pt_z-yv\,\pt_v,\\[4mm]  
Q_2=-zy\,\pt_y+\frac 12\Big(y^2-z^2+v^2\Big)\pt_z-zv\,\pt_v,\earr\right. \qq 
L_3=-y\,\pt_y-z\,\pt_z-v\,\pt_v, 
\eeq
We need 4 extra Killing vectors to get the 10 dimensional $so(4,1)$ Lie algebra for de Sitter metric. 
They are given by
\beq\barr{l}\dst 
C_1=-\frac 1{\tanh\tht}\,\pt_v-\frac 1v\,\pt_{\tht},\\[5mm]\dst 
C_2=\frac 1{\tanh\tht}(v\,\pt_y-y\,\pt_v)-\frac yv\,\pt_{\tht},\qq\quad     
C_3=\frac 1{\tanh\tht}(v\,\pt_z-z\,\pt_v)-\frac zv\ \pt_{\tht},\\[5mm]\dst
C_4=-\frac v{\tanh\tht}(y\,\pt_y+z\,\pt_z)+\frac{(y^2+z^2-v^2)}{2\tanh\tht}\,\pt_v
+\frac{y^2+z^2+v^2}{2v}\,\pt_{\tht}.\earr
\eeq

\section{Elliptic functions: some tools}
There are plenty of books on elliptic function theory, but we used mainly the books by 
Byrd and Friedman \cite{bf} and by Whittaker and Watson \cite{ww}. We use Jacobi rather 
than Weierstrass notation for elliptic functions. Similarly we use earlier Jacobi notation 
for the theta functions which is best adapted to our purposes. They are related to 
the more symmetric notations used in \cite{ww} according to
\[H(v)=\tht_1(w),\quad H_1(v)=\tht_2(w),\quad  \T_1(v)=\tht_3(w),\quad \T(v)=\tht_4(w),
\qq w=\frac{\pi v}{2K}.\]

Let us start from the relation (\ref{deft})
\beq\label{deft2}
\frac{2dt}{\sqrt{3}}=\frac{\rho\,d\rho}{\sqrt{P(\rho)}}.
\eeq
If the quartic polynomial $P(\rho)$ has 2 real roots, and therefore two complex conjugate ones, we will write it
\[P(\rho)=(\rho-a)(\rho-b)[(\rho-b_1)^2+a_1^2],\qq a>b.\]
In this case, the positivity of $\rho$ and $P(\rho)$ requires $\rho\geq a$. One defines
\[A=\sqrt{(a-b_1)^2+a_1^2}\quad > \quad B=\sqrt{(b-b_1)^2+a_1^2},\qq k^2=\frac{(A+B)^2-(a-b)^2}{4AB}<1,\]
where $k^2$ will be the parameter of the elliptic functions involved. Let us define the change of variable
\beq\label{cofv}
{\rm sn}^2\,v=\frac{2B(\rho-a)}{D_+},\qq {\rm cn}^2\,v=\frac{D_-}{D_+},\qq  
{\rm dn}^2\,v=\frac{D_-}{2A(\rho-b)},\eeq
with
\beq
D_{\pm}=A(\rho-b)\pm B(\rho-a)+(a-b)\sqrt{(\rho-b_1)^2+a_1^2},\eeq
and the parameters
\[s_0\equiv {\rm sn}\,v_0=\sqrt{\frac{2B}{A+B+a-b}}<1,\quad 
s_1\equiv {\rm sn}\,v_1=\sqrt{\frac{2B}{A+B-a+b}}>1,\]
for which the reader can check that $\,v_1=K+iK'+v_0$. 

The change of variable (\ref{cofv}) transforms $\,\rho\in\,[a,+\nf)$ into $\,v\in\,[0,v_0)\subset [0,K_0)$.
 The inverse relation is \footnote{From now on we will use the simplified notations 
$s\equiv{\rm sn}\,(v,k^2),\ c\equiv{\rm cn}\,(v,k^2),\ d\equiv{\rm dn}\,(v,k^2)$ as well as 
$s_0={\rm sn}\,v_0,\ s_1={\rm sn}\,v_1$ etc... }
\beq\label{rho}
\rho=\frac{aB\,c^2-bA\,s^2 d^2}{B\,c^2-A\,s^2 d^2}.
\eeq
Using  
\[\barr{c}\dst 
\frac{\rho-a}{a-b}=\frac{A\,s^2 d^2}{B\,c^2-A\,s^2 d^2},\quad  
\frac{\rho-b}{a-b}=\frac{B\,c^2}{B\,c^2-A\,s^2 d^2},\\[5mm]\dst    
\sqrt{(\rho-b_1)^2+a_1^2}=AB\ \frac{d^2-c^2+c^2 d^2}{B\,c^2-A\,s^2 d^2},\earr\]
straightforward computations give
\[\frac{d\rho}{\sqrt{P(\rho)}}=\frac 2{\sqrt{AB}}\ dv.\]
It remains to give the explicit form of $\ga^2=e^{2t}$ as a function of $v$ by 
integrating (\ref{deft2}), which becomes now: 
\beq
\frac{2dt}{\sqrt{3}}=\frac 2{\sqrt{AB}}\frac{aB\,c^2-bA\,s^2\,d^2}{B\,c^2-A\,s^2\,d^2}\,dv.\eeq
The relation
\[\frac{c_0^2}{s^2-s_0^2}=-\frac{c_0}{2s_0d_0}\left(\frac{H'}{H}(v_0-v)+\frac{H'}{H}(v_0+v)
-2\frac{\T'}{\T}(v_0)\right),\]
and a similar one, obtained by the substitution $v_0\to v_1=K+iK'+v_0$:
\[\frac{c_1^2}{s^2-s_1^2}=\frac{c_0}{2s_0d_0}\left(\frac{\T_1'}{\T_1}(v_0-v)+\frac{\T_1'}{\T_1}(v_0+v)
-2\frac{H_1'}{H_1}(v_0)\right),\]
allow us to integrate up to
\beq
\ga^2\equiv e^{2t}=\left(e^{-\xi v}\frac{H(v_0+v)\,\T_1(v_0+v)}{H(v_0-v)\,\T_1(v_0-v)}\right)^{\sqrt{3}},
\qq \xi=2\left(-\frac b{\sqrt{AB}}+\frac{\T'}{\T}(v_0)+\frac{H_1'}{H_1}(v_0)\right).
\eeq
As the reader may notice, in \cite{bf}[p. 135] a different change of variables is given, which 
differs from ours. It is
\[{\rm cn}\, u=\frac{(A-B)\rho-bA+aB}{(A+B)\rho-bA-aB}.\]
As a consequence we get in the metric (\ref{finmet}) the term
\[-\frac 34\,\frac{d\rho^2}{P(\rho)}=-\frac 3{AB}\,\Big(\frac{du}{2}\Big)^2.\]
To avoid the $1/2$ factor we have used a duplication transformation to switch to our 
variable by $u=2v$, having in mind that in the limit $\la\to\,0$ we have $3/AB\to\,1$.

\section{Curvature computations}
Taking the obvious vierbein $e_A$, we define the connection $\om$ and its self-dual components by
\beq
de_A+\om_{AB}\wedge e_B=0,\quad A=0,1,2,3\qq \om^{\pm}_a=\om_{0a}\pm\frac 12\,\eps_{abc}\,\om_{bc},
\quad a,b,c=1,2,3.\eeq
The self-dual components of the curvature follow from
\beq
R_a^{+}=d\om^{+}-\frac 12\,\eps_{abc}\,\om_b^+\wedge\om_c^+,\qq 
R_a^{-}=d\om^{-}+\frac 12\,\eps_{abc}\,\om_b^-\wedge\om_c^-.\eeq
Using the 2-forms
\[\la^{\pm}_a=e_0\wedge e_a\pm\frac 12\,\eps_{abc}\,e_b\wedge e_c,\]
the curvature can be expressed in terms of a triplet of  $\,3\times 3$ matrices
\beq\label{curv}
\left(\barr{c} R_a^+\\ R_a^-\earr\right)=\left(\barr{rr} A_{ab} & B_{ab}\\ ^tB_{ab} & C_{ab}\earr\right)
\left(\barr{c}\la_b^+\\ \la_b^-\earr\right),\qq ^tA=A,\qq ^tC=C.\eeq
Notice that the self-dual components of the Weyl tensor, defined by
\[W^+_a=W^+_{ab}\,\la^+_a,\qq\quad W^-_a=W^-_{ab}\,\la^-_a,\]
are 
\beq
W^+=A-\frac{{\rm tr}\,A}{3}\,{\mb I},\qq\quad W^-=C-\frac{{\rm tr}\,C}{3}\,{\mb I}.\eeq
For the diagonal Bianchi type V metrics considered here the matrices $W^{\pm}$  
have the general structure
\beq
W^{\pm}=\left(\barr{ccc}w_{11} & 0 & 0\\ 0 & w_{22} & \pm\,w_{23}\\ 0 & \pm\,w_{23} & w_{33}\earr\right).\eeq
For the case $\la>0$ we get:
\beq\left\{\barr{l}\dst 
w_{11}=\frac{8\la}{3(1-s^2)^3},\qq w_{23}=-\frac{2\la}{(1-s^2)^2},\\[4mm]\dst
w_{22}=\frac{2\la}{3}\frac{(\sqrt{3}s^3-3\sqrt{3}s-2)}{(1-s^2)^3},\\[4mm]\dst
w_{33}=-\frac{2\la}{3}\frac{(\sqrt{3}s^3-3\sqrt{3}s+2)}{(1-s^2)^3},\earr\right.\eeq
The eigenvalues are all different hence we have a ``Petrov-like" type of the form $\,(I^+,I^-)$.
The conclusions are the same for the case $\la<0$ for which we have
\beq\left\{\barr{l}\dst 
w_{11}=\frac{-8\la s^6}{(3-s^2)^3},\qq w_{23}=\frac{2\la s^4}{(3-s^2)^2},\\[4mm]\dst
w_{22}=\frac{2\la}{3}\frac{s^3(2s^3-9s^2+9)}{(3-s^2)^3},\\[4mm]\dst
w_{33}=\frac{2\la}{3}\frac{s^3(2s^3+9s^2-9)}{(3-s^2)^3}.\earr\right.\eeq

\end{document}